\documentclass[amsfonts, amssymb, amsmath, pra, twocolumn, superscriptaddress, english, floatfix]{revtex4-2}
\usepackage[utf8]{inputenc}
\usepackage{physics}
\usepackage{xcolor}
\usepackage{mathtools}
\usepackage[shortlabels]{enumitem}
\usepackage[colorlinks=true,citecolor=blue,linkcolor=blue,urlcolor=black]{hyperref}
\usepackage[nolist]{acronym}
\usepackage{mathdots}
\usepackage{centernot}
\usepackage{bm}

\newcommand{\DHS}[2]{D_\text{HS}\left(#1, #2\right)}
\newcommand{\EHS}[1]{E_\text{HS}\left(#1\right)}

\newcommand{\cW}{\mathcal{W}}
\newcommand{\cC}{\mathcal{C}}

\newcommand{\GHZ}{\ket{\text{\acs{GHZ}}}}

\begin{document}

\title{A Variational Approach to the Quantum Separability Problem}

\author{Mirko Consiglio}
\email{mirko.consiglio@um.edu.mt}
\affiliation{Department of Physics, University of Malta, Msida MSD 2080, Malta}

\author{Tony J. G. Apollaro}
\affiliation{Department of Physics, University of Malta, Msida MSD 2080, Malta}

\author{Marcin Wie\'sniak}
\affiliation{Institute of Theoretical Physics and Astrophysics, Faculty of Mathematics, Physics, and Informatics, University of Gda\'nsk, 80-308 Gda\'nsk, Poland}
\affiliation{International Centre for Theory of Quantum Technologies, University of Gda\'nsk, 80-308 Gda\'nsk, Poland}

\begin{abstract}
    We present the \ac{VSV}, which is a novel \ac{VQA} that determines the \ac{CSS} of an arbitrary quantum state with respect to the \ac{HSD}. We first assess the performance of the \ac{VSV} by investigating the convergence of the optimization procedure for \ac{GHZ} states of up to seven qubits, using both statevector and shot-based simulations. We also numerically determine the \ac{CSS} of \ac{X-MEMS}, and subsequently use the results of the algorithm to surmise the analytical form of the aforementioned \ac{CSS}. Our results indicate that current \ac{NISQ} devices may be useful in addressing the $NP$-hard full separability problem using the \ac{VSV}, due to the shallow quantum circuit imposed by employing the destructive \textsc{SWAP} test to evaluate the \ac{HSD}. The \ac{VSV} may also possibly lead to the characterization of multipartite quantum states, once the algorithm is adapted and improved to obtain the \ac{k-CSS} of a multipartite entangled state.
\end{abstract}
\date{\today}

\maketitle

\begin{acronym}
\acro{CPTP}{completely-positive trace-preserving}
\acro{CSS}{closest separable state}
\acro{GHZ}{Greenberger--Horne--Zeilinger \acroextra{(state)}}
\acro{GME}{genuine multipartite entanglement}
\acro{GSA}{generalized simulated annealing}
\acro{hc}[h.c.]{Hermitian conjugate}
\acro{HSD}{Hilbert--Schmidt distance}
\acro{HSE}{Hilbert--Schmidt entanglement}
\acro{k-CSS}[$k$-CSS]{closest $k$-separable state}
\acro{KKT}{Karush--Kuhn--Tucker \acroextra{(conditions)}}
\acro{LOCC}{local operations and classical communication}
\acro{NFT}{Nakanishi--Fujii--Todo \acroextra{(optimiser)}}
\acro{NISQ}{noisy intermediate-scale quantum}
\acro{PPT}{positive partial transpose}
\acro{QGA}{quantum Gilbert algorithm}
\acro{SLOCC}{stochastic local operations and classical communication}
\acro{SLSQP}{sequential least squares programming \acroextra{(optimiser)}}
\acro{VSV}{variational separability verifier}
\acro{VQA}{variational quantum algorithm}
\acro{X-MEMS}[$X$-MEMS]{maximally-entangled mixed $X$-states}
\end{acronym}

\acresetall
\section{Introduction} \label{sec:intro}

Entanglement is the principal defining feature of quantum mechanics~\cite{Einstein1935}, and a quantitative description of the phenomenon started with Bell's inequalities~\cite{Bell1964}. Different measures of entanglement quantify distinct resources, although most require the fulfillment of specific conditions to be considered ``good'' entanglement measures. These conditions can be defined as a set of axioms, such as the nullification of the measure for separable states, invariance under local unitary operations, and monotonicity under \ac{LOCC}~\cite{Vedral1997}. To this end, many entanglement measures have been defined, such as distillable entanglement, entanglement cost, and the entanglement of formation~\cite{Plenio2007, Eltschka2014}. The difficulty in calculating these measures for generic states lies in their non-closed form, given that they generally include the computation of a minimum over a large Hilbert space, which is typically intractable as the dimension of the system scales. On the other hand, the problem of measuring entanglement is more approachable for pure states, given that even for different entanglement classes, that is under \ac{SLOCC}~\cite{Eltschka2014}, many different measures of entanglement can be computed. 

The simplest system that has been completely characterized is a two-qubit one. An interesting property for two qubits is that only two \ac{SLOCC} classes exist, one which contains the only type of entanglement possible, i.e. bipartite entanglement, and the other consists of fully separable states, i.e. non-entangled states. The relevant literature for three- and four-qubit pure states, as well as the general multipartite case can be found in Refs.~\cite{Coffman2000, Cunha2019, Guhne2010, Dur2000},~\cite{Regula2014, Verstraete2002, Gour2010, Ghahi2016, Osterloh2016}, and~\cite{Guo2020, Eisert2001, Love2006, Guhne2010, Bengtsson2016} respectively. Only highly specific mixed states are known to have closed forms of multipartite entanglement measures, such as combinations of \ac{GHZ} and W states~\cite{Eltschka2008}, \ac{GHZ}-symmetric states~\cite{Eltschka2012}, and \ac{X-MEMS}~\cite{Agarwal2013}, among a few others. It is also known that any arbitrary teleportation protocol~\cite{Bennett1993}, using an arbitrary multipartite entanglement channel requires an entanglement monotone termed localizable concurrence~\cite{Popp2005}, as a quantum resource~\cite{Consiglio2021}.

We can also relax our specificity of requiring an exact quantifier of entanglement, and focus on whether a particular state is certified to be entangled. This is known as the quantum separability problem, and is in fact deemed to be $NP$-hard, even for the bipartite case~\cite{Gurvits2004, Ioannou2007, Gharibian2010}. Consequently, verifying a state is \textit{fully} separable requires that the state is separable with respect to all bipartitions, which implies that the full separability problem is at least as hard as the bipartite separability problem. As a result, we look towards \acp{VQA}, which incorporate hybrid quantum--classical computation aimed at harnessing the power of \ac{NISQ} computers~\cite{Preskill2018}, to solve challenging computational problems.

To this end, we describe a novel \ac{VQA}, the \ac{VSV}, capable of determining the \ac{CSS} of an arbitrary quantum state (assuming it can be prepared on a quantum device), with respect to the \ac{HSD}~\cite{Bengtsson2006}. In addition, the \ac{HSD} induces an entanglement measure, denoted as the \ac{HSE}~\cite{Witte1999, Bengtsson2006}. The \ac{HSE} is still up for debate whether it is an entanglement monotone~\cite{Ozawa2000}, however, it could in some cases: quantify the amount of entanglement present in specific states; behave as a separability witness; or else provide useful constructions of entanglement witnesses~\cite{Pandya2020}.

We will also briefly mention the prospect of extending the \ac{VSV} to find the \ac{k-CSS}, while for the time being, the algorithm is designed to find the closest \textit{fully} separable state, i.e. the 1-\ac{CSS}, which for the sake of brevity we refer to as the \ac{CSS}. Although we investigate only qubit systems, the \ac{VSV} can be adapted to find the \ac{CSS} for qudit systems, however, this requires an encoding of the $d$-level system onto a 2-level system, as well as adjustments to the gates needed to generate ($k$-)separable qudit states.

The paper is organized as follows: Sec.~\ref{sec:HSD} consists of definitions for the \ac{HSD}, the induced \ac{HSE} measure, and the \ac{CSS}, along with a discussion on why we use the \ac{HSD} as our metric for the \ac{VSV}. In Sec.~\ref{sec:framework}, we introduce the framework of the \ac{VSV}, specifically how minimizing the \ac{HSD} over the set of fully separable states, with respect to a state $\rho$, leads to the \ac{CSS} of $\rho$ and its \ac{HSE}. Since the computation of the \ac{HSD} requires the calculation of state overlaps, we discuss how this is tackled on a quantum device using the destructive \textsc{SWAP} test~\cite{Garcia-Escartin2013, Cincio2018}. We then introduce the variational form of the algorithm and explain the corresponding bilevel optimization~\cite{Dempe2002} approach. Lastly, we discuss some of the computational complexity aspects of the \ac{VSV}, such as the scaling of the number of parameters with respect to the dimension of the system. In Sec.~\ref{sec:VSV_results} we present the results of the \ac{VSV} applied to $n$-qubit \ac{GHZ} states, ranging from two to seven qubits with statevector optimization, and from two to five qubits with shot-based optimization. We also apply the \ac{VSV} to two- and three-qubit \ac{X-MEMS}. Since the analytical form of the \ac{CSS} for $n$-qubit \ac{X-MEMS} is not known, we employed a technique similar to the one in Ref.~\cite{Verstraete2002} to surmise the analytical form of the \ac{CSS} in Appendix~\ref{app:A}. Furthermore, we show that the \ac{GME} concurrence~\cite{Eltschka2014} and \ac{HSE} are not related in the case of three-qubit $X$-States~\cite{Agarwal2013}. Finally, in Sec.~\ref{sec:conc}, we summarize the results of the paper, and acknowledge the inspiration of the \ac{VSV}: the \ac{QGA}~\cite{Brierley2016, Shang2018, Wiesniak2020, Pandya2020}, which is further discussed in Appendix~\ref{app:B}.

\section{The Hilbert--Schmidt Distance as a Measure of Entanglement} \label{sec:HSD}

The \ac{HSD} between two quantum states $\DHS{\rho}{\sigma}$, is defined as
\begin{equation}
    \DHS{\rho}{\sigma} \equiv \Tr{\left( \rho - \sigma \right)^2},
    \label{eq:HSD}
\end{equation}
which is a non-monotonic Riemannian metric~\cite{Bengtsson2006}. Using this definition, an entanglement measure can be induced by the \ac{HSD}~\cite{Witte1999, Bengtsson2006}, denoted as the \ac{HSE},
\begin{equation}
    \EHS{\rho} \equiv \underset{\sigma \in \cC}{\min}~\DHS{\rho}{\sigma},
    \label{eq:EHS_min}
\end{equation}
with the \ac{CSS} defined as
\begin{equation}
    \rho_\text{CSS} \equiv \underset{\sigma \in \cC}{\arg\min}~\DHS{\rho}{\sigma}.
    \label{eq:CSS_min}
\end{equation}
where $\cC$ is the convex set of states for some degree $k$ of separability. To define it concretely, a state $\rho$ is said to be $k$-separable if it can be written as a convex sum of $k$-separable states, that is
\begin{equation}
    \rho = \sum_i p_i \ketbra{\Psi_i^k},
\end{equation}
where a $k$-separable pure state can be written as a tensor product of $k$-local states
\begin{equation}
    \ket{\Psi_i^k} = \bigotimes_{j=1}^k \ket{\psi_j},
\end{equation}
such that $\ket{\psi_j}$ are $k$-separable states on subsets of the $n$-qubit parties~\cite{Shang2018}. More specifically, states are called biseparable for $k = 2$, triseparable for $k = 3$, up to fully separable for $k = n$. Although the \ac{k-CSS}~\eqref{eq:CSS_min} (and corresponding \ac{HSE}~\eqref{eq:EHS_min}) can be distinctly defined for all convex sets of $k$-separable states, as stated in Sec.~\ref{sec:intro}, we designed the \ac{VSV} to find the \ac{CSS} (and corresponding \ac{HSE}) of entangled states with respect to the set of \textit{fully} separable states.

It was proven in Ref.~\cite{Ozawa2000} that the \ac{HSD} is not non-increasing under \ac{CPTP} maps, posing a principal question of whether the \ac{HSE} is a good entanglement measure. Nevertheless, the \ac{HSD}, and the corresponding \ac{HSE}, are still useful quantities to investigate, since they are utilized in generalized Bell inequalities~\cite{Bertlmann2002, Silva2022}, while also providing insight into the geometry of entangled states~\cite{Bengtsson2006, Streltsov2010}. The \ac{HSD} is also relatively straightforward to evaluate on a quantum device, since it can be decomposed as
\begin{equation}
    \Tr{\left( \rho - \sigma \right)^2} = \Tr{\rho^2} + \Tr{\sigma^2} - 2\Tr{\rho\sigma},
    \label{eq:HSD_decomp}
\end{equation}
coupled with using quantum primitives such as the destructive \textsc{SWAP} test to measure the overlap and purities of $\rho$ and $\sigma$.

On the other hand, while it was shown in Ref.~\cite{Streltsov2010} that distance measures, such as the Bures measure of entanglement~\cite{Bengtsson2006}, are directly related to the fidelity~\cite{Uhlmann1976} of a \ac{CSS}, the fidelity is a harder quantity to evaluate on both a classical and quantum device~\cite{Witte1999, Brun2004, Garcia-Escartin2013, Bartkiewicz2013, Cincio2018, Cerezo2020} when compared with the simplicity of the \ac{HSD}.

It is also interesting to mention that the \ac{HSD} is being used for tackling the quantum low-rank approximation problem~\cite{Ezzell2022a}, with applications in principal component analysis and preparing arbitrary mixed states on quantum computers~\cite{Ezzell2022b}.

\section{Framework of the Algorithm}  \label{sec:framework}

In our procedure, we designate $\rho$ as being the test state, and $\sigma$ as our trial state(s), which are the states that iteratively approach the \ac{CSS}. As a result, we choose to prepare separable states on a quantum computer in the following way:
\begin{equation}
    \sigma(\bm{p}, \bm{\theta}, \bm{\phi}) = \sum_{i=1}^{s} p_i \ketbra{\psi (\bm{\theta}_i, \bm{\phi}_i)}{\psi(\bm{\theta}_i, \bm{\phi}_i)},
    \label{eq:CSS}
\end{equation}
where
\begin{align}
    \bm{p} &= \left(
    \begin{array}{c}
        p_1 \\
        p_2 \\
        \vdots \\
        p_s \\
    \end{array}
    \right),~\sum\limits_{i=1}^s p_i = 1,~p_i \in [0, 1], \nonumber \\[1ex]
    \bm{\theta} &=
    \left( 
    \begin{array}{c}
        \bm{\theta}_1 \\
        \bm{\theta}_2 \\
        \vdots \\
        \bm{\theta}_s \\
    \end{array}
    \right) =
    \left( 
    \begin{array}{cccc}
        \theta_{11} & \theta_{12} & \cdots & \theta_{1n} \\
        \theta_{21} & \theta_{22} & \cdots & \theta_{2n} \\
        \vdots & \vdots & \ddots & \vdots \\
        \theta_{s1} & \theta_{s2} & \cdots & \theta_{sn} \\
    \end{array} 
    \right),~\theta_{ij} \in [0, 2\pi), \nonumber \\[1ex]
    \bm{\phi} &= 
    \left(
    \begin{array}{c}
        \bm{\phi}_1 \\
        \bm{\phi}_2 \\
        \vdots \\
        \bm{\phi}_s \\
    \end{array}
    \right) =
    \left( 
    \begin{array}{cccc}
        \phi_{11} & \phi_{12} & \cdots & \phi_{1n} \\
        \phi_{21} & \phi_{22} & \cdots & \phi_{2n} \\
        \vdots & \vdots & \ddots & \vdots \\
        \phi_{s1} & \phi_{s2} & \cdots & \phi_{sn} \\
    \end{array} 
    \right),~\phi_{ij} \in [0, 2\pi), \nonumber
\end{align}
totaling $s(2n + 1)$ parameters, where $n$ is the number of qubits and $s$ is the number of separable pure states needed to generate the trial state. Due to Carath\'eodory's theorem, we require $s \leq d^2$ separable pure states~\cite{Horodecki1997, Vedral1998, Streltsov2010}, where $d = 2^n$ is the dimension of the Hilbert space. However, in our simulations we found that $s = d$ suffices to find the \ac{CSS} of all our test states. $\left\{ p_i,\ket{\psi(\bm{\theta}_i, \bm{\phi}_i)} \right\}$ represent the ensemble of separable pure states, which can be decomposed as
\begin{equation}
    \ket{\psi(\bm{\theta}_i, \bm{\phi}_i)} = \bigotimes_{j=1}^n \left( \cos\left(\theta_{ij}\right)\ket{0_j} + e^{\imath\phi_{ij}}\sin\left(\theta_{ij}\right)\ket{1_j} \right).
\end{equation}
A separable pure state can thus be generated on a quantum computer by applying a set of one-qubit gates to the all-zero state:
\begin{equation}
    \ket{\psi(\bm{\theta}_i, \bm{\phi}_i)} = \bigotimes_{j=1}^n U(\theta_{ij}, \phi_{ij}) \ket{0}_j,
    \label{eq:unitary}
\end{equation}
where
\begin{equation}
    U(\theta_{ij}, \phi_{ij}) = \left( \begin{array}{cc}
        \cos\left(\theta_{ij}\right) & -\sin\left(\theta_{ij}\right) \\[1ex]
        e^{\imath\phi_{ij}}\sin\left(\theta_{ij}\right) & e^{\imath\phi_{ij}}\cos\left(\theta_{ij}\right)
    \end{array} \right),
\end{equation}
which is equivalent to applying an $R_y(\theta_{ij})$ gate followed by an $R_z(\phi_{ij})$ gate.

\subsection{Measuring the Hilbert--Schmidt Distance} \label{sec:HSD_measure}

Preparing a mixed state on a quantum computer, requires, in general, twice as many qubits when compared to the size of the state~\cite{Benenti2009}. This, coupled with the issue of preparing an arbitrary separable state, leads us to find an alternative method for computing the overlap of the trial states with the test state. Our solution is to individually supply separable (non-orthogonal) pure states of our separable mixed state on the quantum computer, evaluate the individual overlaps using the destructive \textsc{SWAP} test, and then classically mix them during the optimization procedure. The benefit of this method is that we do not require the direct preparation of the \ac{CSS} on the quantum computer, as we are only interested in the computation of the \ac{HSD}.

Suppose, we can decompose the trial state $\sigma$ as in Eq.~\eqref{eq:CSS}, then we can evaluate the overlap between $\rho$ and $\sigma$ as
\begin{align}
    \Tr{\rho\sigma} &= \Tr{\rho \sum_{i=1}^s p_i \ketbra{\psi_i}{\psi_i}} \nonumber \\
    &= \sum_{i=1}^s p_i \Tr{\rho \ketbra{\psi_i}{\psi_i}} \nonumber \\
    &= \sum_{i=1}^s p_i \ev{\rho}{\psi_i}.
    \label{eq:overlap}
\end{align}
If we want to compute the purity of $\sigma$, then
\begin{align}
    \Tr{\sigma^2} &= \Tr{\sum_{i=1}^s p_i \ketbra{\psi_i}{\psi_i} \sum_{j=1}^s p_j
    \ketbra{\psi_j}{\psi_j}} \nonumber \\
    &= \sum_{i,j=1}^s p_i p_j \Tr{\ket{\psi_i}\braket{\psi_i}{\psi_j}\bra{\psi_j}} \nonumber \\
    &= \sum_{i,j=1}^s p_i p_j |\braket{\psi_i}{\psi_j}|^2 \nonumber \\
    &= \sum_{i=1}^s p_i^2 + \sum_{i \neq j}^s p_i p_j |\braket{\psi_i}{\psi_j}|^2 \nonumber \\
    &= \sum_{i=1}^s p_i^2 + 2\sum_{i<j}^s p_i p_j |\braket{\psi_i}{\psi_j}|^2.
    \label{eq:purity}
\end{align}
The purity of $\rho$ is trivially obtained assuming we can prepare it directly on a quantum computer; given that it is our test state.

\subsection{Measuring Overlaps on a Quantum Computer} \label{sec:SWAP}

\begin{figure}[t]
    \centering
    \includegraphics[width=0.3\textwidth]{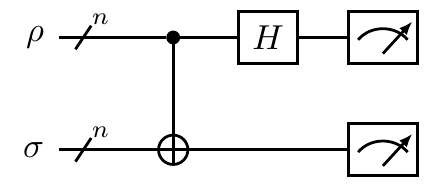}
    \caption[Destructive \textsc{SWAP} test for determining the overlap between two quantum states]{Destructive \textsc{SWAP} test for determining the overlap between two quantum states. The bundled \textsc{CNOT} gate here actually represents \textsc{CNOT} gates between each pair of qubits of $\rho$ and $\sigma$, with the bundled Hadamard gate equating to performing a Hadamard gate on each qubit of $\rho$. This effectively reduces to a Bell basis transformation for each pair of qubits of $\rho$ and $\sigma$.}
    \label{fig:destructive_swap_test}
\end{figure}

Given Eqs~\eqref{eq:overlap} and~\eqref{eq:purity}, we require subroutines in the \ac{VSV} capable of measuring the purity and overlap of quantum states. Specifically, one can utilize the destructive \textsc{SWAP} test~\cite{Garcia-Escartin2013, Cincio2018} to calculate these quantities. The concept of this procedure stems from the fact that measuring in the Bell basis determines the amount of correlations present between two systems, and it can be shown that it is equivalent to the non-destructive \textsc{SWAP} test~\cite{Buhrman2001}. Fig.~\ref{fig:destructive_swap_test} shows the quantum circuit for performing the destructive \textsc{SWAP} test.

By placing a \textsc{CNOT} gate followed by a Hadamard gate on the control qubit, each pair of qubits of $\rho$ and $\sigma$ are measured in the Bell basis. Following this, each pair of measurement results are post-processed using the vector $\vec{c} = (1, 1, 1, -1)$, which corresponds to summing the probabilities of getting 00, 01 and 10, and subtracting the probability of getting 11. This is equivalent to measuring the expectation value of a \textsc{CZ} operator, resulting in obtaining the overlap $\Tr{\rho\sigma}$, as long as the qubits are rearranged as $R_1S_1R_2S_2\dots R_nS_n$, where $R_i$ and $S_i$ denote the subsystems of $\rho$ and $\sigma$, respectively. This results in a linear scaling in post-processing, given that we do not directly compute $\vec{c}\cdot\vec{p}$, where $\vec{p}$ is the probability vector, but rather binning the paired measurement outcomes into a $+1$ and $-1$ bin, and then averaging the results~\cite{Cincio2018}. The purity of a state can be similarly obtained by supplying two copies of the state as inputs to the test. The one-qubit gates necessary to generate the separable pure states of Eq.~\eqref{eq:unitary}, coupled with the two-depth circuit needed for the destructive \textsc{SWAP} test, results in a noticeably shallow circuit for the \ac{VSV}.

\subsection{Variational Optimization}

Eqs.~\eqref{eq:overlap} and~\eqref{eq:purity} give us the means to compute the \ac{HSD} as in Eq.~\eqref{eq:HSD_decomp}, and by combining these with Eqs.~\eqref{eq:EHS_min} and~\eqref{eq:CSS_min}, we can devise a \ac{VQA} that is able to compute the \ac{CSS} and the corresponding \ac{HSE}.

The optimizer in this scenario is tasked with providing angles $\bm{\theta}$ and $\bm{\phi}$ to prepare the separable pure states, and probabilities $\bm{p}$ to classically mix them during post-processing, to generate a separable state $\sigma(\bm{p}, \bm{\theta}, \bm{\phi})$. One can notice that the quantum computer is only tasked with computing $\Tr{\rho^2}$, $\ev{\rho}{\psi(\bm{\theta}_i, \bm{\phi}_i)}$, and $|\braket{\psi(\bm{\theta}_i, \bm{\phi}_i)}{\psi(\bm{\theta}_j, \bm{\phi}_j)}|^2$ $\forall~i,j \in [s]$. The probabilities $\bm{p}$ are only incorporated classically when computing the final results for calculating the purity $\Tr{\sigma(\bm{p}, \bm{\theta}, \bm{\phi})^2}$, and the overlap $\Tr{\rho \sigma(\bm{p}, \bm{\theta}, \bm{\phi})}$. As a result, we split our optimization routine into a bilevel system, which is essentially a nested optimization routine~\cite{Dempe2002}.

The structure of the bilevel system is as follows: at the beginning of every iteration, an upper level optimizer selects the parameters $\bm{\theta}$ and $\bm{\phi}$, and proceeds to call the quantum computer to compute the overlaps in Eqs.~\eqref{eq:overlap} and \eqref{eq:purity}. The upper level optimizer then launches a lower level optimizer to obtain the parameters $\bm{p}$ to minimize Eq.~\eqref{eq:HSD_decomp}. We thus obtain (near-)optimal parameters of the minimization of the cost function~\eqref{eq:HSD},
\begin{equation}
    \{\bm{p^*}, \bm{\theta^*}, \bm{\phi^*}\} \approx \underset{\bm{p}, \bm{\theta}, \bm{\phi}}{\arg\min}~\DHS{\rho}{\sigma(\bm{p}, \bm{\theta}, \bm{\phi})},
\end{equation}
such that the \ac{CSS} of $\rho$ would then be
\begin{equation}
    \rho_\text{CSS} \approx \sigma(\bm{p^*}, \bm{\theta^*}, \bm{\phi^*}),
\end{equation}
from which we can calculate the \ac{HSE} of $\rho$ as
\begin{equation}
    \EHS{\rho} \approx \DHS{\rho}{\sigma(\bm{p^*}, \bm{\theta^*}, \bm{\phi^*})}.
\end{equation}

\begin{figure*}[t]
    \centering
    \includegraphics[width=0.7\textwidth]{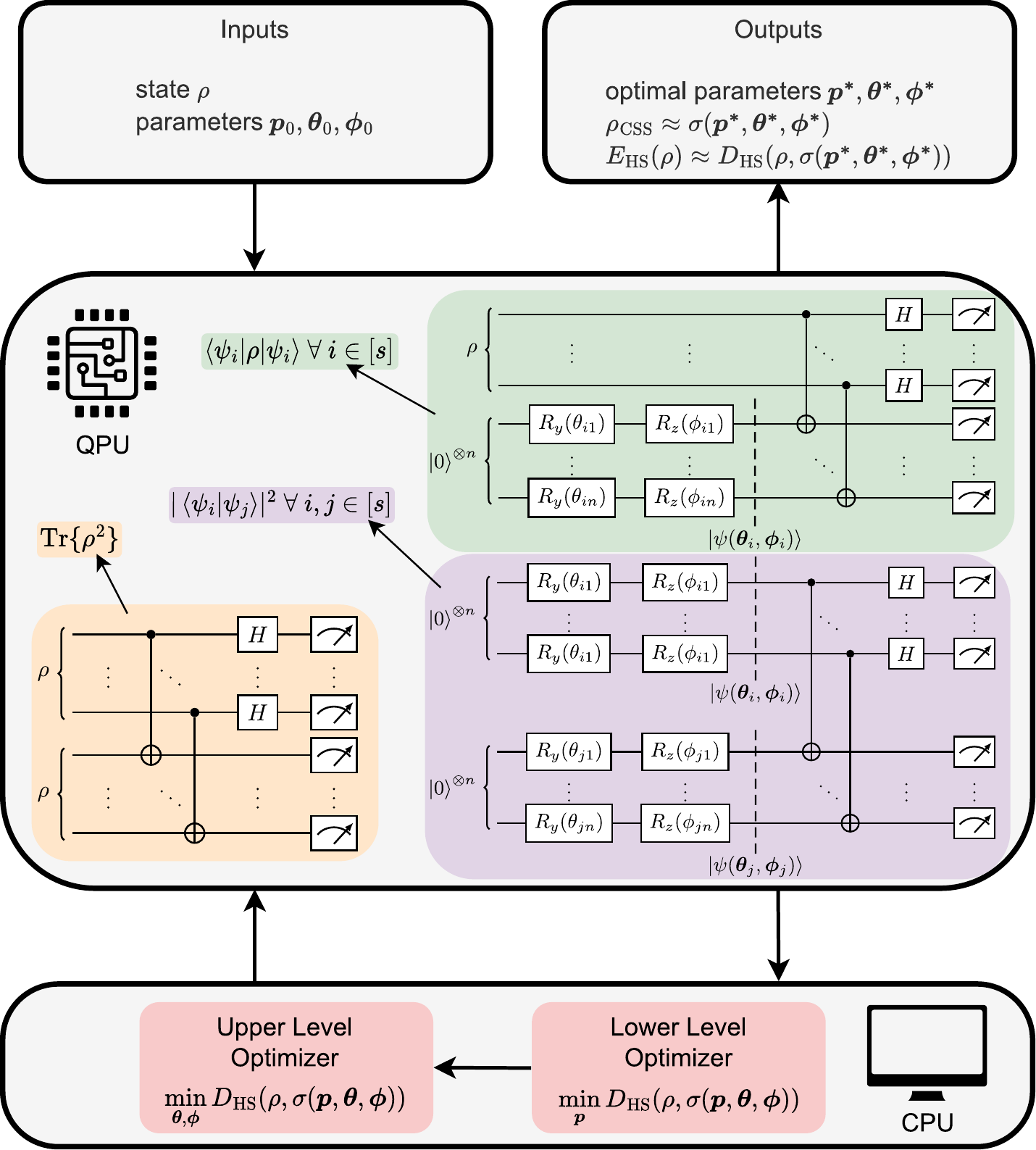}
    \caption{Diagrammatic representation of the \ac{VSV}. The inputs to the \ac{VSV} are the test state $\rho$, and initial parameters $\bm{\rho}_0, \bm{\theta}_0, \bm{\phi}_0$. The \ac{VSV} then computes the orange (bottom-left QPU) circuit once, to compute $\Tr{\rho^2}$, with the green (top-right QPU) and purple (bottom-right QPU) circuits evaluated to compute overlaps $\ev{\rho}{\psi_i}$ and $\braket{\psi_i}{\psi_j}~\forall~i,j \in [s]$, which are used in Eqs.~\eqref{eq:overlap} and~\eqref{eq:purity}, respectively. The red (bottom CPU) section is the classical feedback loop computed by the CPU, consisting of first minimizing the parameters $\bm{p}$, followed by proposing new parameters $\bm{\theta}$ and $\bm{\phi}$ to evaluate on the QPU. After a sufficient number of iterations, the \ac{VSV} outputs optimal parameters $\bm{\rho^*}, \bm{\theta^*}, \bm{\phi^*}$, along with the computed \ac{CSS} and corresponding \ac{HSE}.}
    \label{fig:diagram}
\end{figure*}

A diagrammatic representation of the \ac{VSV} algorithm is shown in Fig.~\ref{fig:diagram}.

\subsection{Complexity Analysis} \label{sec:complexity}

The scalability of the \ac{VSV}, similar to other hybrid quantum--classical algorithms, is characterized from both the classical and quantum point of view. The algorithm prepares the test state $\rho$ (which may be unknown, but reproducible) and trial states $\sigma$ on a quantum computer, which provides an exponential memory reduction when compared with purely classical algorithms. We also assume $\rho$ can be efficiently prepared on a quantum device: either via direct input, for example, from another quantum system; or else using quantum state preparation methods~\cite{Araujo2021}. Aside from preparing $\rho$, the algorithm requires $n$ \textsc{CNOT} gates and $n$ Hadamard gates for carrying out the destructive \textsc{SWAP} test, which scales linearly with the number of qubits, and uses a constant circuit depth of two, assuming the quantum computer has a ladder-like connectivity.

In the current implementation of the \ac{VSV}, the number of parameters scales as $s(2n + 1)$ for determining the closest \textit{fully} separable state of $\rho$, where $s$ is the number of (non-orthogonal) components representing the trial state $\sigma$. Unless $\rho$ is separable, we found that the \ac{CSS} of $\rho$ being full rank. Thus, we require at least $s \geq d$, although we found that in all of our simulations $s = d$ suffices to find the \ac{CSS} of an entangled state. This implies that we have an exponential scaling in the number of parameters needed to determine the \ac{CSS} via this method. This may be alleviated if one finds a way to directly prepare the \ac{CSS} as a separable mixed state on the quantum device, which will (generally) require doubling the number of qubits~\cite{Benenti2009}. On the other hand, the number of parameters necessary to define the \ac{CSS} may be reduced such that it scales sub-exponentially with the number of qubits. It should also be noted that finding the \ac{k-CSS} would require more complex ans\"atze, which are more likely to require a larger set of parameters to represent $k$-separable states.

\section{Results} \label{sec:VSV_results}

The results involving the \ac{VSV} are presented in this section for both statevector and shot-based simulations, with all initial points being generated randomly. Statevector simulations refer to instances where we have perfect information about the quantum states (that is access to the vector of the state), which is used to benchmark and test the algorithm. On the other hand, shot-based simulations refer to the cases where we sample the state obtained at the end of each circuit to obtain $N$ bit strings, where $N$ is denoted as the number of shots.

\begin{figure*}[t]
    \centering
    \includegraphics[width=0.875\textwidth]{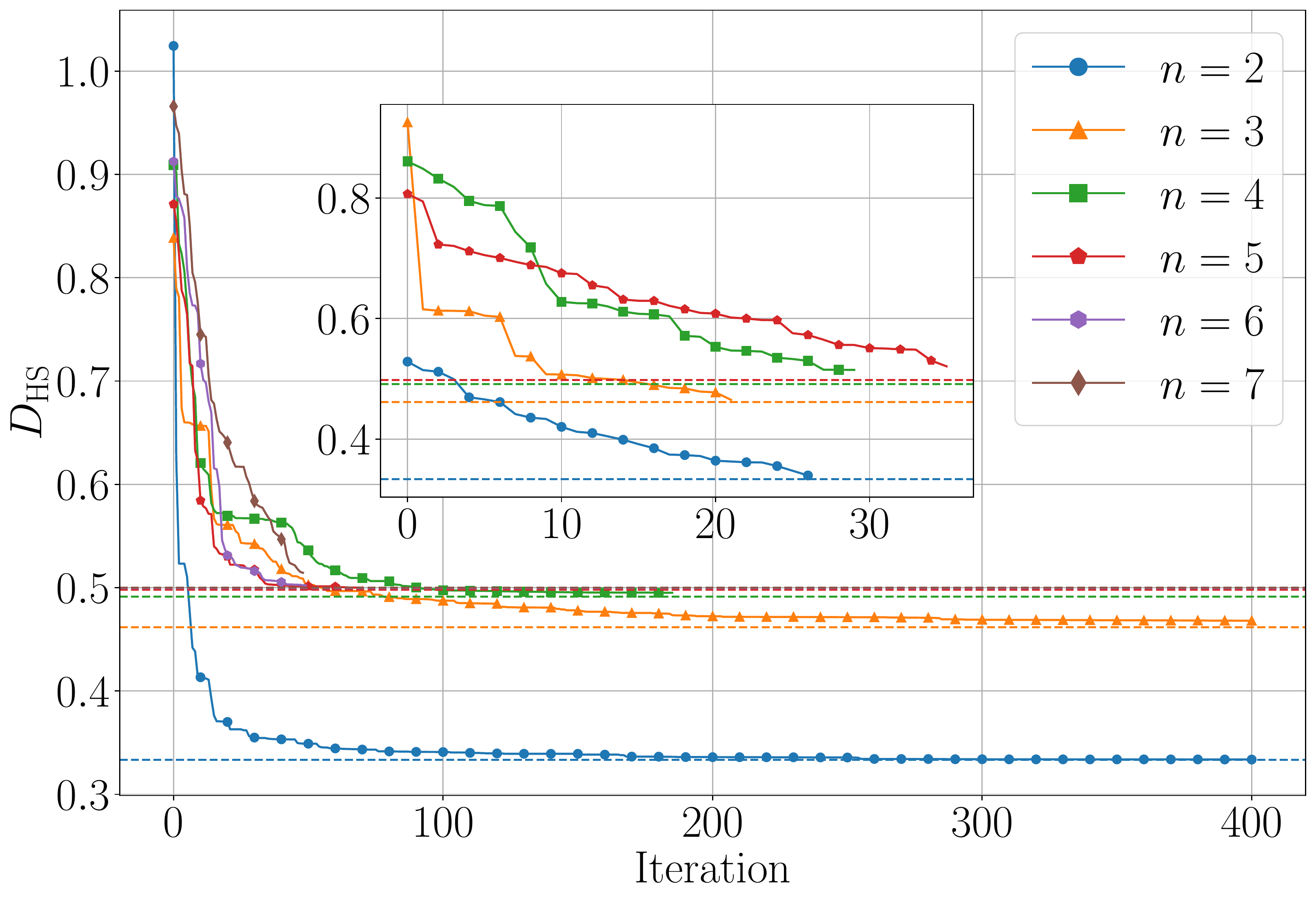}
    \caption{Plot of the convergence of \ac{HSD} with respect to the number of iterations for the statevector optimization for \ac{GHZ} states. The solid lines represent the convergence for two to seven qubits (bottom-up). The dashed lines represent the analytical \ac{HSE} of the corresponding number of qubits, that is $\EHS{\GHZ_n}$. The inset figure represents the shot-based optimization convergence with 8192 shots for two to five qubits (bottom-up). Note that there are less iterations for the shot-based cases (although same number of function evaluations), since it is harder for the \ac{GSA} optimizer to find a strictly lower \ac{HSE} at each function evaluation.}
    \label{fig:GHZ_plot}
\end{figure*}

A plot of the \ac{HSD} convergence using statevector optimization for \ac{GHZ} states, ranging from two to seven qubits, is shown in Fig.~\ref{fig:GHZ_plot} --- with an inset similarly showing two to five qubits but using shot-based optimization with 8192 shots. The optimizer used in this instance is the \ac{GSA} algorithm~\cite{Tsallis1996, Xiang1997} from the \texttt{SciPy} library~\cite{dual_annealing}, for upper parameter ($\bm{\theta}$, $\bm{\phi}$) optimization, with the \ac{SLSQP} algorithm~\cite{Kraft1988} for the lower parameter ($\bm{p}$) optimizer. The \ac{GHZ} states are specifically chosen so as to demonstrate the performance of the \ac{VSV}, and also since the \ac{CSS} is analytically known, as given in Ref.~\cite{Pandya2020}.

\ac{GHZ} states are the basis of numerous quantum communication protocols, such as quantum secret sharing~\cite{Hillery1999}, and quantum communication complexity reduction~\cite{Ho2022}, beside being the experimental workhorse for more refined tests of quantum non-locality with respect to two-qubit Bell states~\cite{Pan2000}. \ac{GHZ}-diagonal states may inherit their non-classical features from the former~\cite{Chen2012}, since they are typically a consequence of local damping channels acting on individual qubits. Thus, investigating the \ac{CSS} may provide us with a better understanding of the optimal strategies in the presence of noise, as well as insight into the amount of error that is admissible in these protocols, in order to retain a quantum advantage.

\begin{figure*}[t]
    \centering
    \begin{minipage}{0.46\textwidth}
        \centering
        \includegraphics[width=\textwidth]{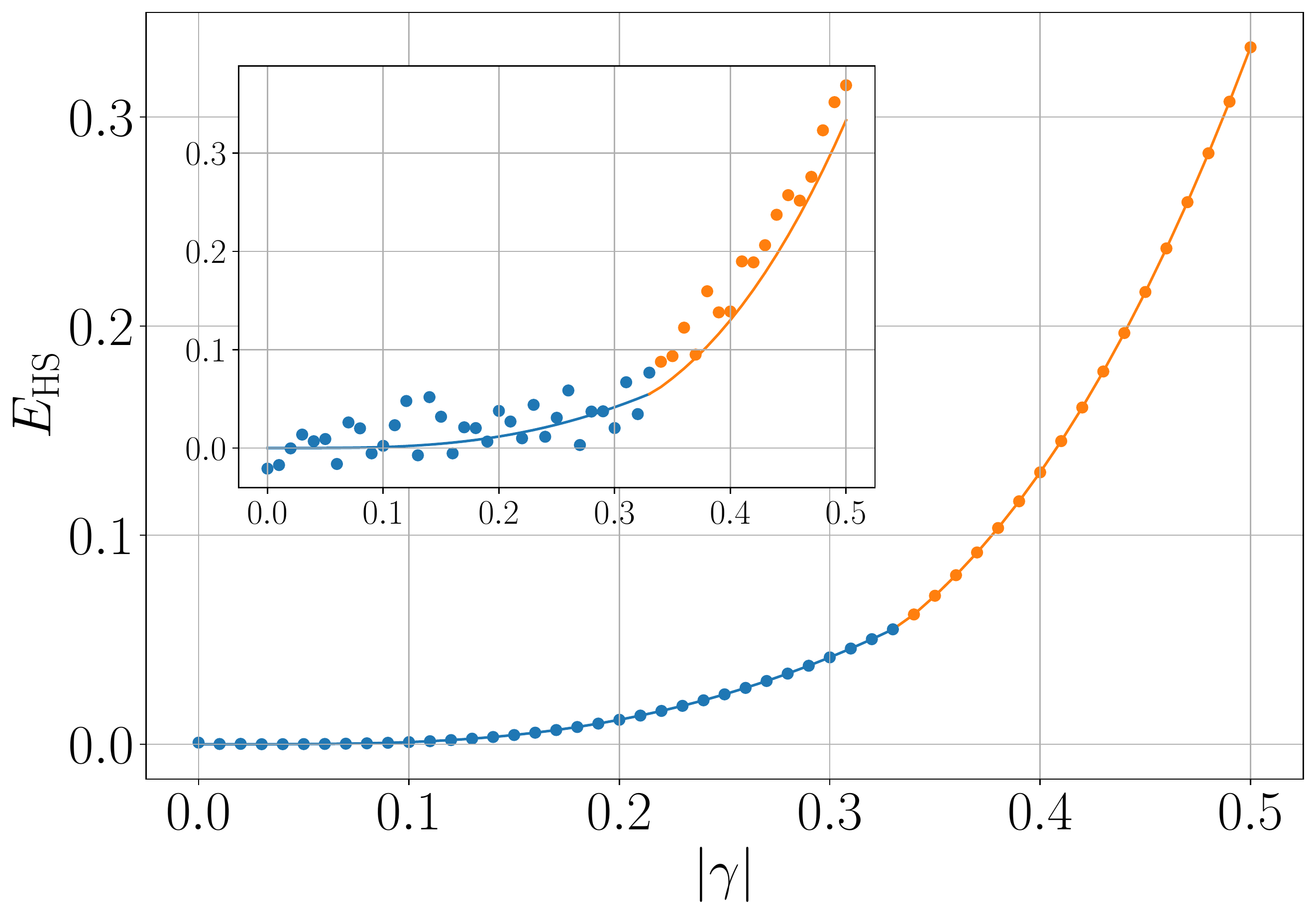}
        \caption{Plot of the results of the statevector optimization applied to two-qubit \ac{X-MEMS}. The solid lines represent the analytical \ac{HSE} as a function of $|\gamma|$. The left blue segment represents the range between $0 \leq |\gamma| \leq \frac{1}{3}$, while the right orange segment represents the range between $\frac{1}{3} \leq |\gamma| \leq \frac{1}{2}$. The inset figure represents the data acquired with shot-based optimization using 8192 shots.}
        \label{fig:X-MEMS_2}
    \end{minipage}
    \hspace{0.04\textwidth}
    \begin{minipage}{0.46\textwidth}
        \centering
        \includegraphics[width=\textwidth]{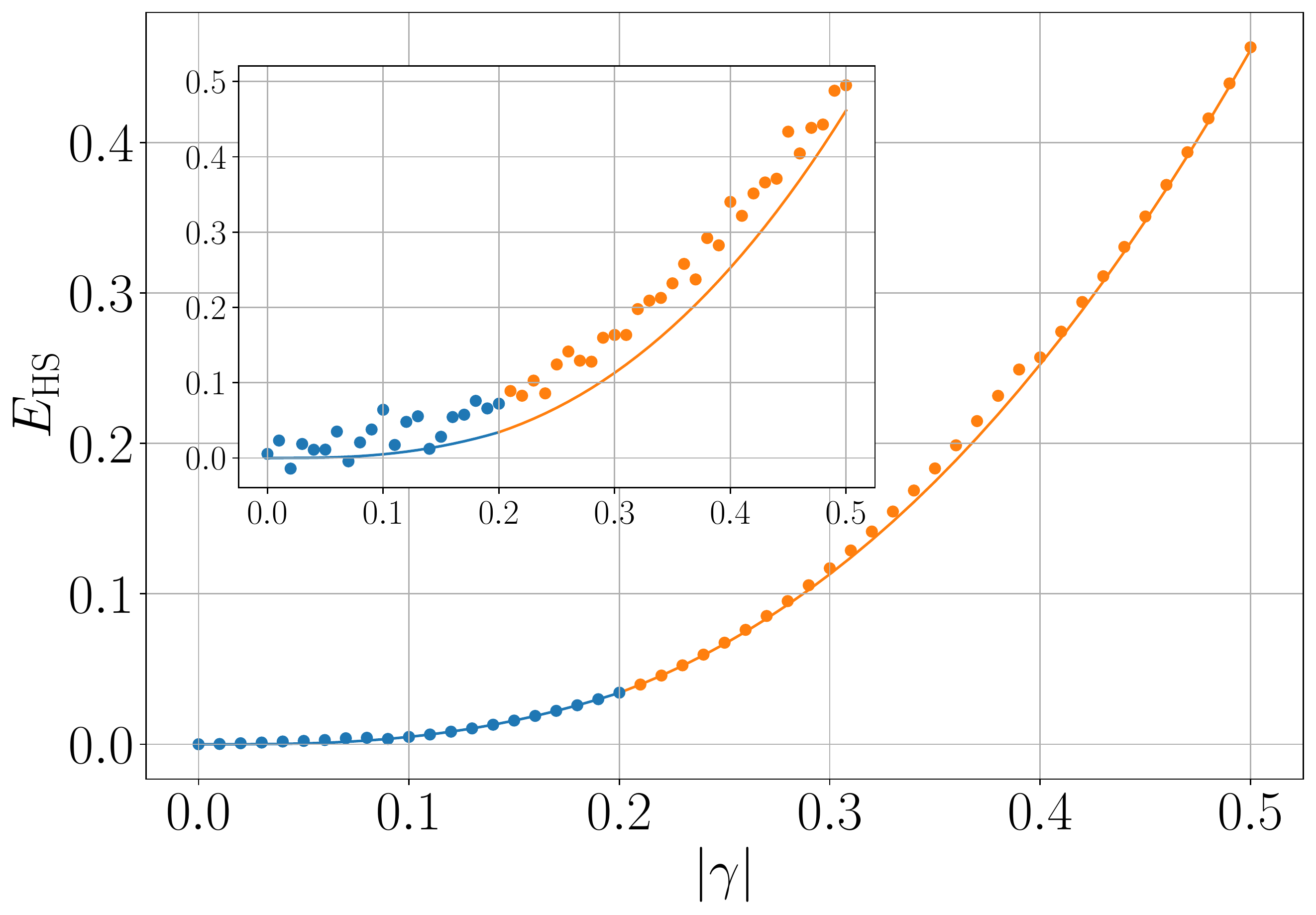}
        \caption{Plot of the results of the statevector optimization applied to three-qubit \ac{X-MEMS}. The solid lines represent the analytical \ac{HSE} as a function of $|\gamma|$. The left blue segment represents the range between $0 \leq |\gamma| \leq \frac{1}{5}$, while the right orange segment represents the range between $\frac{1}{5} \leq |\gamma| \leq \frac{1}{2}$. The inset figure represents the data acquired with shot-based optimization using 8192 shots.}
        \label{fig:X-MEMS_3}
    \end{minipage}
\end{figure*}

The following application of the \ac{VSV} is on two- and three-qubit \ac{X-MEMS}, which are $X$-states that have the maximal amount of multipartite entanglement for a given linear entropy~\cite{Agarwal2013}. These states are specifically chosen since the \ac{GME} concurrence~\cite{Eltschka2014}, is known exactly for $X$-states. The \ac{GME} concurrence is an extension of the bipartite concurrence, which is a measure related to the entanglement of formation~\cite{Hill1997, Wootters1998}, and in fact the \ac{GME} concurrence reduces to the bipartite concurrence for two-qubit states. The intention is to relate the \ac{HSE} with the \ac{GME} concurrence, possibly enabling the \ac{HSE} to behave as an entanglement monotone for \ac{X-MEMS}, albeit it does not satisfy the conditions for one in general~\cite{Bengtsson2006}. Figs.~\ref{fig:X-MEMS_2} and~\ref{fig:X-MEMS_3} show the analytical \ac{HSE} along with the values determined by the \ac{VSV}, for two- and three-qubit \ac{X-MEMS}. In the case of the inset figures, the \ac{NFT} optimizer~\cite{Nakanishi2020} was used as an upper level optimizer, since it offered significant improvement over \ac{GSA} when evaluating the overlaps using shots rather than statevector calculations.

By inspection of the \ac{CSS} determined by the \ac{VSV}, we were able to surmise the form of the \ac{CSS} for $n$-qubit \ac{X-MEMS}, which is
\begin{equation}
    \left( 
    \begin{array}{cccccccc}
        \frac{a}{2} & & & & & & & d \\
        & \frac{b}{N - 1} & & & & & 0 & \\
        & & \ddots & & & \iddots & & \\
        & & & \frac{b}{N - 1} & 0 & & & \\
        & & & 0 & \frac{1 - a - b}{N - 1} & & & \\
        & & \iddots & & & \ddots & & \\
        & 0 & & & & & \frac{1 - a - b}{N - 1} & \\
        d^* & & & & & & & \frac{a}{2} \\
    \end{array} 
    \right),
\end{equation}
where $N = 2^{n-1}$, $a, b \in [0, 1]$, $a + b \leq 1$, and $|d| \in \left[0, \frac{a}{2}\right]$, to ensure a valid density matrix. The parameters $a$, $b$ and $d$ can be directly determined by using the \texttt{Mathematica} code on \texttt{GitHub}~\cite{VSV_code}. We found that for $n \geq 3$, $a$ is a root of a quartic equation including $\gamma$, with $b$ and $d$ depending on both $a$ and $\gamma$. More details can be found in Appendix~\ref{app:A}.

Similar to \ac{GHZ} states and their \acp{CSS} (which are also $X$-states), the $X$-state itself is robust against local damping channels~\cite{Yu2007, Hashemi2012}. This is due to the fact that decoherence effects that introduce additional non-$X$ elements, to a density matrix, cannot decrease the degree of entanglement~\cite{Agarwal2013}. As a consequence, the $X$ part of a density matrix provides a lower bound on the \ac{GME} present in the state~\cite{Ma2011}. Furthermore, the possibility of an arbitrary initial pure state decohering into an $X$-state was studied in Ref.~\cite{Quesada2012}.

In the case of two-qubit states, the \ac{HSE} and the concurrence is known to be equal for a restricted set of states~\cite{Mundarain2007, Ganardi2022}. With our \ac{VSV}, we have numerical evidence supporting the conjecture that the \ac{HSE} of two distinct, arbitrary, two-qubit states, $\rho$ and $\sigma$, is equal if, and only if, the concurrence of $\rho$ and $\sigma$ is also equal, that is
\begin{equation}
    C(\rho) = C(\sigma) \iff \EHS{\rho} = \EHS{\sigma},
\end{equation}
and similarly, that the \ac{HSE} of $\rho$ is greater than that of $\sigma$ if, and only if, the concurrence of $\rho$ is also greater than that of $\sigma$, that is
\begin{equation}
    C(\rho) > C(\sigma) \iff \EHS{\rho} > \EHS{\sigma}.
\end{equation}
Following this, we tested whether similar statements hold when comparing $\ac{GME}$ concurrence and \ac{HSE} for $(n \geq 3)$-qubit $X$-states. All of the simulations we carried out pointed towards the null hypothesis in both cases, meaning that equality does not hold:
\begin{equation}
    C_\text{GME}(\rho) = C_\text{GME}(\sigma) \centernot\iff \EHS{\rho} = \EHS{\sigma},
\end{equation}
nor does ordered inequality:
\begin{equation}
    C_\text{GME}(\rho) > C_\text{GME}(\sigma) \centernot\iff \EHS{\rho} > \EHS{\sigma}.
\end{equation}
This is not surprising, since the \ac{GME} concurrence characterizes $n$-qubit entanglement. On the other hand, the \ac{HSE} seems to describe the departure of a state from a fully separable one, where in the case of three-qubits, may also consist of biseparable states which are not detected by \ac{GME} concurrence.

\acresetall
\section{Conclusion} \label{sec:conc}

The \ac{VSV} is a novel \ac{VQA} that finds the \ac{CSS} of arbitrary quantum states, with respect to the \ac{HSD}, and obtain the \ac{HSE}. The \ac{VSV} has been applied to $n$-qubit \ac{GHZ} states, showing the convergence of the optimizer. It has also been applied to investigate \ac{X-MEMS}, producing a relation between the \ac{GME} concurrence and \ac{HSE}, as well as helping deduce the analytical form of the \ac{CSS} for $n$-qubit \ac{X-MEMS}. In this respect, the \ac{VSV} can be useful for shedding light on possible analytical forms of \acp{CSS} for arbitrary states. Further simulations carried out on two-qubit states demonstrated that while equality (or ordering) of the concurrence of two states is present if, and only if, equality (or ordering) of the \ac{HSE} is also present, this does not hold in general for the \ac{GME} concurrence and \ac{HSE} in the case of $X$-states with three or more qubits.

If the calculated \ac{HSE} of an entangled state is equal to zero, then we trivially acquire $\rho_\text{CSS} = \rho$, which immediately implies that our test state $\rho$ is fully separable, and hence, is not entangled. In either case, if one knows the \ac{CSS}, then an entanglement witness~\cite{Horodecki1996, Doherty2004, Guhne2009} can be defined as $\cW \equiv \rho - \rho_\text{CSS}$. In general however, one does not exactly acquire the \ac{CSS}, but rather a close approximation to it, say, $\sigma$. In this case, the optimal entanglement witness would be of the form
$\cW \equiv \rho - \sigma - \max_{\ket{\psi} \in \cC} \ev{(\rho - \sigma)}{\psi}$~\cite{Pandya2020}.

The strength of the \ac{VSV} as a potential \ac{NISQ} algorithm stems from its simplicity. While the algorithm requires in general many calls to the quantum computer for evaluating overlaps, the destructive \textsc{SWAP} test, coupled with the ease of preparing separable states through one-qubit gates, results in a three-depth circuit which is remarkably tractable on a \ac{NISQ} device. It should be noted that the entire trial state(s) can be directly prepared on the quantum computer by utilizing (up to $n$) ancillary qubits, which are then traced out after performing the relevant unitary gates~\cite{Benenti2009}, leading to less overlap computations for the cost of more qubits, as discussed in Sec.~\ref{sec:framework}. It is also noteworthy to mention that the evaluation of state overlaps could be improved by looking towards alternative methods, such as in Refs.~\cite{Flammia2011, Elben2019, Elben2020, Fanizza2020, Liu2022}.

Future work would involve enhancing the \ac{VSV} to find the \ac{k-CSS} of an arbitrary quantum state, which is not straightforward, since even the closest biseparable state for three-qubit states is significantly challenging to implement. The issue lies in properly expressing a biseparable state in a variational form, since the number of pure state bipartitions scales as $2^{n-1} - 1$ for an $n$-qubit state. The set of all biseparable states is generated by the convex hull of all pure state bipartitions. This, coupled with the fact that an $n$-qubit state can have rank up to $2^n$ makes the extension of the \ac{VSV} to determine the \ac{k-CSS} extremely non-trivial. The question of finding the \ac{k-CSS}, along with improvements to the performance of the \ac{VSV}, are left as open problems.

Apart from the \ac{VSV}, there are alternative approaches that aim to find the \ac{CSS} of arbitrary states using other methods, such as neural networks~\cite{Girardin2022} and an adaptive polytope approximation~\cite{Ohst2022}. However, work that inspired the variational approach to the problem at hand is the so-called \ac{QGA}~\cite{Brierley2016, Shang2018, Wiesniak2020, Pandya2020}. The implementation of the \ac{QGA} on a quantum device and related discussion can be found in Appendix~\ref{app:B}.

The \texttt{Python} code for running the \ac{VSV} simulations, using the package \texttt{Qulacs}~\cite{Suzuki2021}, and the \texttt{Mathematica} code for determining the \ac{CSS} of \ac{X-MEMS}, can be found on \texttt{GitHub}~\cite{VSV_code}.

\section*{Acknowledgments}

MC acknowledges funding by the Tertiary Education Scholarships Scheme and the MCST Research Excellence Programme 2022 on the QVAQT project at the University of Malta. TJGA acknowledges funding by the SEA-EU Alliance, European University of the Seas, Research Seed Fund ``STATED''. MW acknowledges partial support by NCN Grant No. 2017/26/E/ST2/01008 and the Foundation for Polish Science (IRAP project, ICTQT, Contract No. 2018/MAB/5, co-financed by EU within Smart Growth Operational Programme).

\appendix

\section{Deriving the CSS for \textit{X}-MEMS} \label{app:A}

The general form of \ac{X-MEMS} is given by
\begin{equation}
    \tilde{X} = \left( 
    \begin{array}{cccccccc}
        f(\gamma) & & & & & & & \gamma \\
        & g(\gamma) & & & & & 0 & \\
        & & \ddots & & & \iddots & & \\
        & & & g(\gamma) & 0 & & & \\
        & & & 0 & 0 & & & \\
        & & \iddots & & & \ddots & & \\
        & 0 & & & & & 0 & \\
        \gamma^* & & & & & & & g(\gamma) \\
    \end{array} 
    \right),
\end{equation}
where

\begin{subequations}
\label{eq:fg}
\begin{align}
    f(\gamma) &= 
    \begin{cases}
        \frac{1}{N+1} & 0 \leq |\gamma| \leq \frac{1}{N+1}, \\
        |\gamma| & \frac{1}{N+1} \leq |\gamma| \leq \frac{1}{2},
    \end{cases} \\
    g(\gamma) &= 
    \begin{cases}
        \frac{1}{N+1} & 0 \leq |\gamma| \leq \frac{1}{N+1}, \\
        \frac{1 - 2|\gamma|}{N-1} & \frac{1}{N+1} \leq |\gamma| \leq \frac{1}{2},
    \end{cases}
\end{align}
\end{subequations}
and $N = 2^{n - 1}$. The \ac{GME} concurrence was shown to be equal to $2|\gamma|$ for \ac{X-MEMS}~\cite{Agarwal2013}.

\subsection{Two-qubit \textit{X}-MEMS}

The analytical form of the \ac{CSS} for two-qubit \ac{X-MEMS} was derived using analytical optimization with the \ac{KKT} conditions~\cite{Tabak1971}, since the numerical results by the \ac{VSV} hinted towards a \ac{CSS} of the form
\begin{equation}
    \left(
    \begin{array}{cccc}
    \frac{a}{2} & 0 & 0 & d \\
    0 & b & 0 & 0 \\
    0 & 0 & 1 - a - b & 0 \\
    d^* & 0 & 0 & \frac{a}{2}
    \end{array}
    \right),
\end{equation}
where $a, b \in [0, 1]$, $a + b \leq 1$, and $|d| \in \left[0, \frac{a}{2}\right]$, to ensure a valid density matrix. For this state to be separable, then the concurrence must be equal to zero, meaning that
\begin{equation}
    |d|^2 \leq b(1 - a - b).
    \label{eq:no_conc}
\end{equation}
Optimizing this problem results in parameters 
\begin{subequations}
\begin{align}
    a &= \frac{1}{9}\left(
    7 - \sqrt{1 + 36 |\gamma|^2}
    \right), \\
    b &= \frac{1 + 12 |\gamma|^2+\sqrt{1 + 36 |\gamma|^2}}{6 \sqrt{1 + 36|\gamma|^2}}, \\
    d &= \frac{\gamma}{3}  \left(1 + \frac{2}{\sqrt{1 + 36 |\gamma|^2}}\right),
\end{align}
\end{subequations}
with an \ac{HSE} of
\begin{equation}
    \EHS{\tilde{X}} = \frac{2}{27} \left(1 + 18 |\gamma|^2-\sqrt{1 + 36 |\gamma|^2}\right),
    \label{eq:HSE_1}
\end{equation}
for $0 \leq |\gamma| \leq \frac{1}{3}$, and
\begin{subequations}
\begin{align}
    a &= \frac{1}{3} \left(1 + 4|\gamma| - \sqrt{1 - 4|\gamma| + 8|\gamma|^2}\right), \\
    b &= \frac{1}{6} \left(3 - 6|\gamma| + \frac{3 - 12|\gamma| + 16 |\gamma|^2}{\sqrt{1 - 4|\gamma| + 8|\gamma|^2}}\right), \\
    d &= \frac{\gamma  \left(2 - 4|\gamma| + \sqrt{1 - 4|\gamma| + 8|\gamma|^2}\right)}{3 \sqrt{1 - 4|\gamma| + 8|\gamma|^2}},
\end{align}
with an \ac{HSE} of
\begin{align}
    \EHS{\tilde{X}} &= \frac{2}{3} \left(1 - 4|\gamma| + 6|\gamma|^2 \right. \nonumber \\ 
    & \hspace{0.85cm} + \left. (2|\gamma| - 1)\sqrt{1 - 4|\gamma| + 8|\gamma|^2} \right),
    \label{eq:HSE_2}
\end{align}
\end{subequations}
for $\frac{1}{3} \leq |\gamma| \leq \frac{1}{2}$. 

\subsection{Three-qubit \textit{X}-MEMS}

The analytical form of the \ac{CSS} for three-qubit \ac{X-MEMS} was derived using analytical optimization with the \ac{KKT} conditions~\cite{Tabak1971}, since the numerical results by the \ac{VSV} hinted towards a \ac{CSS} of the form
\begin{equation}
    \left(
    \begin{array}{cccccccc}
    \frac{a}{2} & 0 & 0 & 0 & 0 & 0 & 0 & d \\
    0 & \frac{b}{3} & 0 & 0 & 0 & 0 & 0 & 0 \\
    0 & 0 & \frac{b}{3} & 0 & 0 & 0 & 0 & 0 \\
    0 & 0 & 0 & \frac{b}{3} & 0 & 0 & 0 & 0 \\
    0 & 0 & 0 & 0 & \frac{1 - a - b}{3} & 0 & 0 & 0 \\
    0 & 0 & 0 & 0 & 0 & \frac{1 - a - b}{3} & 0 & 0 \\
    0 & 0 & 0 & 0 & 0 & 0 & \frac{1 - a - b}{3} & 0 \\
    d^* & 0 & 0 & 0 & 0 & 0 & 0 & \frac{a}{2}
    \end{array}
    \right),
    \label{eq:three_qubit_CSS}
\end{equation}
where similarly, $a, b \in [0, 1]$, $a + b \leq 1$, and $|d| \in \left[0, \frac{a}{2}\right]$, to ensure a valid density matrix. Since we require the state to be separable, we first need to certify that the \ac{GME} concurrence is zero, which results in the same condition as in~\eqref{eq:no_conc}. However, the \ac{GME} concurrence only assures us that the state does not have tripartite entanglement in this scenario, yet we still need to ensure that the state has no bipartite entanglement. It can be immediately seen from the density matrix~\eqref{eq:three_qubit_CSS} that the concurrence between pairs of individual qubits is zero --- since the partial trace with respect to each qubit results in a diagonal matrix. 

What is left to check is the entanglement between each qubit and the rest, and so, we look to the negativity~\cite{Zyczkowski1998}, although the negativity does not capture bound entangled states~\cite{Horodecki1996, Peres1996, Horodecki1998}. However, we conjecture that the state in~\eqref{eq:three_qubit_CSS} is not bound entangled, since the numerical \ac{HSE}, determined by the \ac{VSV}, coincides with the \ac{HSE} of~\eqref{eq:three_qubit_CSS}, as shown in Fig.~\ref{fig:X-MEMS_3}. Due to the uniqueness of the \ac{CSS}~\cite{Pandya2020}, this implies that the fully separable state determined by the \ac{VSV}, is exactly the same state as~\eqref{eq:three_qubit_CSS}, meaning that it is fully separable.

The partial transpose of~\eqref{eq:three_qubit_CSS} with respect to each qubit is equivalent, and the negativity is non-zero if the partial transpose has negative eigenvalues. Now, the partial transpose of~\eqref{eq:three_qubit_CSS} only has one eigenvalue which can be negative, for some parameters $a$, $b$ and $d$, which is
\begin{equation}
    \frac{1}{6}\left( 1 - a - \sqrt{(1 - a - 2b)^2 + 36|d|^2} \right),
\end{equation}
meaning we need to ascertain that the above condition is non-negative, for~\eqref{eq:three_qubit_CSS} to be separable. This condition can instead be rewritten and shown that it is equivalent to
\begin{equation}
    |d|^2 \leq \frac{b(1 - a - b)}{9},
    \label{eq:no_neg}
\end{equation}
which interestingly encompasses condition~\eqref{eq:no_conc} as well, resulting in an analytical optimization problem using the \ac{KKT} conditions with only constraint~\eqref{eq:no_neg} to insure separability. For the sake of brevity, we shall not provide the parameters $a$, $b$ and $d$, as well as the \ac{HSE} here, since they are in terms of a root of a quartic equation and so the explicit equations would be too cumbersome to present. Nevertheless, they can be reproduced using the \texttt{Mathematica} code available on \texttt{GitHub}~\cite{VSV_code}.

\subsection{\textit{n}-qubit \textit{X}-MEMS}

While we have not explicitly derived the \ac{CSS} of $n$-qubit \ac{X-MEMS} in general, we conjecture that it is of the form
\begin{equation}
    \left( 
    \begin{array}{cccccccc}
        \frac{a}{2} & & & & & & & d \\
        & \frac{b}{N - 1} & & & & & 0 & \\
        & & \ddots & & & \iddots & & \\
        & & & \frac{b}{N - 1} & 0 & & & \\
        & & & 0 & \frac{1 - a - b}{N - 1} & & & \\
        & & \iddots & & & \ddots & & \\
        & 0 & & & & & \frac{1 - a - b}{N - 1} & \\
        d^* & & & & & & & \frac{a}{2} \\
    \end{array} 
    \right),
\end{equation}
where $N = 2^{n-1}$, $a, b \in [0, 1]$, $a + b \leq 1$, and $|d| \in \left[0, \frac{a}{2}\right]$, to ensure a valid density matrix. This form is also invariant under the symmetries imposed by \ac{X-MEMS}, which is a necessary condition for the \ac{CSS} as shown in Ref.~\cite{Pandya2020}. The symmetries of the \ac{CSS} consist of any qubit permutations from the set $\{2, \dots, n\}$, and the condition that the all-zero element must be equal to the all-one element. The parameter $a$ in the \ac{CSS} seems to in general be a root of a quartic equation, with the parameters $b$ and $d$ depending on both $a$ and $\gamma$. The condition for zero negativity also encompasses the cases for $(k \geq 3)$-separability, which is equivalent to
\begin{equation}
    |d|^2 \leq \frac{b(1 - a - b)}{(N - 1)^2},
\end{equation}
presumably related to the hierarchy of mixed state entanglement~\cite{Eltschka2014}.

\begin{figure}[t]
    \centering
    \includegraphics[width=0.46\textwidth]{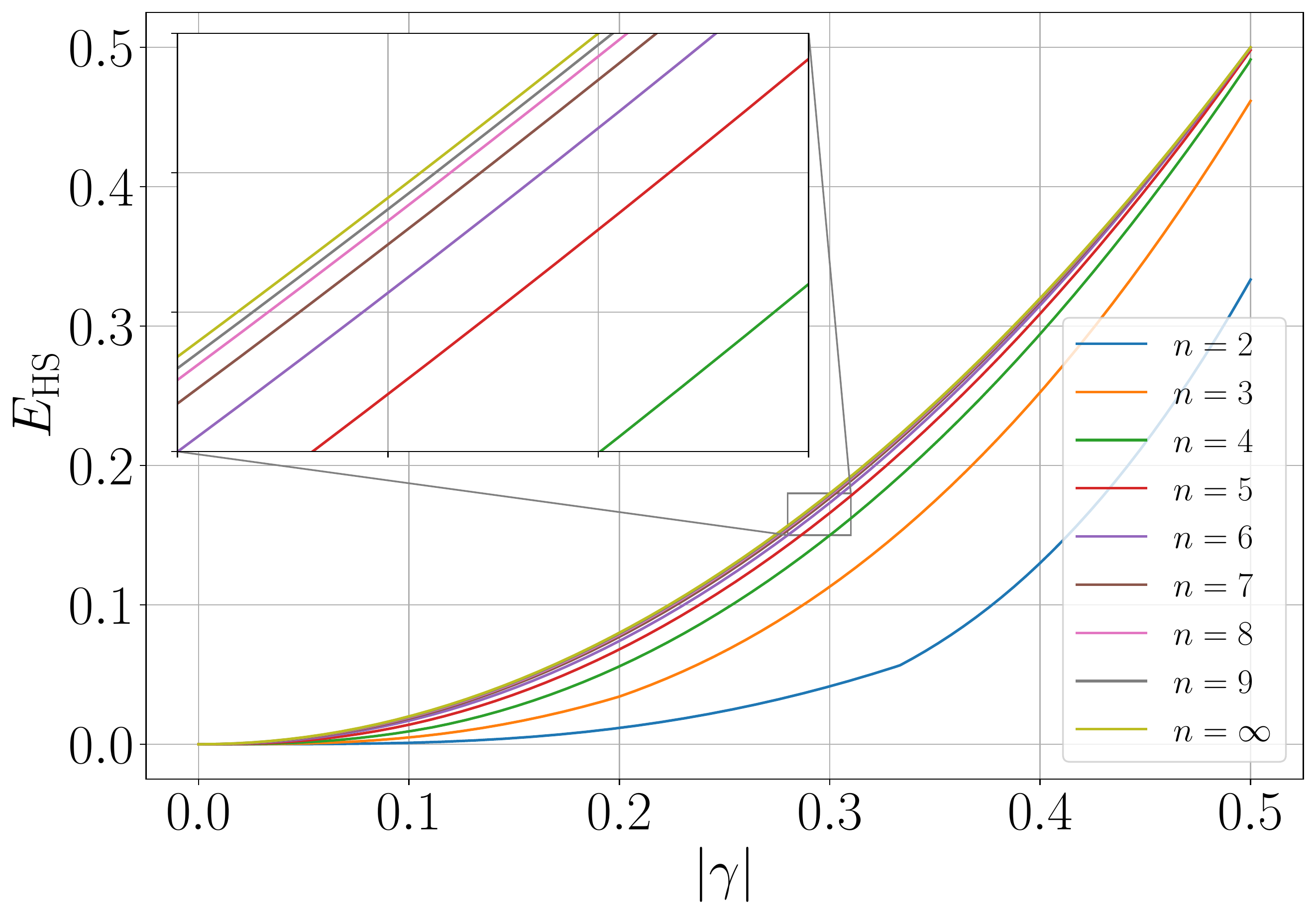}
    \caption{Analytical \ac{HSE} of two- to nine-qubit \ac{X-MEMS} as a function of $|\gamma|$ (bottom-up). The infinite-qubit (top-most line) case represents the analytical upper bound of $2|\gamma|^2$ for the \ac{HSE} of \ac{X-MEMS}.}
    \label{fig:analytical_X-MEMS}
\end{figure}

Fig.~\ref{fig:analytical_X-MEMS} shows the analytical \ac{HSE} of two- to nine-qubit \ac{X-MEMS} as a function of $|\gamma|$. Each function for the \ac{HSE} is piece-wise continuous, with the domains divided at the point $|\gamma| = \frac{1}{N + 1}$, as is similarly given by the definition of the \ac{X-MEMS} in Eqs.~\eqref{eq:fg}. The \acp{CSS} possess an \ac{HSE} of zero at $|\gamma| = 0$ and an \ac{HSE} of
\begin{equation}
    \EHS{\GHZ_n} = \frac{2^n-2}{2^{n + 1} + 2^{3-n} - 4},
\end{equation}
at $|\gamma| = \frac{1}{2}$, since the $n$-qubit \ac{X-MEMS} correspond to the $n$-qubit \ac{GHZ} states at that point~\cite{Pandya2020}. It is also interesting to note the limit of the functions in Fig.~\ref{fig:analytical_X-MEMS}, corresponding to the infinite-qubit $\ac{X-MEMS}$, which is equal to
\begin{equation}
    \lim\limits_{n \rightarrow \infty} \EHS{\tilde{X}_n} = 2|\gamma|^2,
\end{equation}
meaning that the \ac{HSE} for \ac{X-MEMS} is bounded above.

\section{Comparison with the Quantum Gilbert Algorithm} \label{app:B}

The \ac{QGA} is the quantum analogue of Gilbert's algorithm~\cite{Gilbert1966}, which is utilized to determine the distance between a given point and a convex set, according to the Hilbert-Schmidt norm. The \ac{QGA} has insofar only been implemented on classical computers to test for membership of states in different \ac{SLOCC} classes~\cite{Brierley2016, Shang2018, Wiesniak2020, Pandya2020}.

It is only natural to attempt to extend the \ac{QGA} to operate on a quantum computer, owing to many reasons. The first of which is the storage of quantum mixed state data, which is known to increase exponentially as $4^n$, with respect to the number of qubits $n$. The second is the input of arbitrary mixed states into a quantum computer, created via open quantum systems as an example, where these reasons also apply to the \ac{VQA} counterpart. The issue in simply extending the \ac{QGA} to be utilized on a quantum computer, in order to improve the performance over its classical counterpart, is due to the nature of the algorithm itself, requiring possibly, an infinite number of state mixtures to the trial state. The \ac{QGA} relies on computing overlaps of different trial states, similar to the \ac{VSV}, however, the main difference is that the \ac{VSV} keeps a constant number of states in memory. On the other hand, the \ac{QGA}, at each successful iteration of the algorithm adds one state to memory, increasing the number of future overlap calculations. This means that more calls to the quantum computer are required for subsequent iterations of the algorithm. In general if we suppose we require one million trial states for a three-qubit state to acquire a reasonable \ac{CSS}, which is exactly the amount of states taken by the \ac{QGA}, as given in Ref.~\cite{Pandya2020}, and we assume that the number of successes, $c_s$, is proportional to the number of trial states, $c_t$, by $c_s \propto c_t^{0.44127}$, then the number of overlaps is equal to
\begin{equation}
    \sum_{i = 1}^{10^6} \sum_{j = 1}^{c_s^{(i)}} j = \sum_{i = 1}^{10^6} \frac{c_s^{(i)}\left(c_s^{(i)} + 1\right)}{2},
\end{equation}
where $c_s^{(i)}$ is the number of successes at iteration $i$, which starts at $1$ for $i = 1$, and increases by $1$ every $\frac{c_t}{c_s} \approx \frac{c_t}{c_t^{0.44127}} = c_t^{0.55873}$ times, which for one million trial states is approximately every $2251$ trial states. However, in the beginning, almost every trial state will correspond to a success, due to the initial state on average lying farther away from the actual \ac{CSS}. As the number of successes increases, and the state moves closer to the border of the set of separable states, nearer to the \ac{CSS}, it will be harder to find a closer separable state. But, even if we take it to be split evenly, then this results to be around $3.3\times10^{10}$ calls to the quantum computer. On the other hand, the \ac{VSV} calls the quantum computer $5\times10^3$ times to obtain a \ac{CSS} comparable to the \ac{QGA}, which would equate to more than $6\times10^6$ calls for the \ac{QGA} for each call of the \ac{VSV}.

In either case, we shall describe the direct implementation of the \ac{QGA} on a quantum device. Given an initial guess in the form of a pure product state $\sigma_0 = \ketbra{\psi_0}{\psi_0}$, after the $n^\text{th}$ success, the new \ac{CSS} $\rho_n$ to the test state $\rho$ is given iteratively as
\begin{equation}
    \rho_{n} = p_n \rho_{n-1} + (1 - p_n)\sigma_n,
    \label{eq:iter}
\end{equation}
which equates to
\begin{equation}
    \rho_{n} = \sum_{i=0}^{n} \left(\prod_{j=i+1}^{n} p_j\right) \left(1-p_i\right) \sigma_i,
\end{equation}
where $\sigma_i = \ketbra{\psi_i}{\psi_i}$ is a pure product state and $p_i \in (0, 1)$, such that
\begin{equation}
    \sum_{i=0}^{n} \left(\prod_{j=i+1}^{n} p_j\right) \left(1-p_i\right) = 1.
\end{equation}
Note that this definition entails that $\rho_0 \equiv \sigma_0$. The implementation of a \ac{QGA} can be thus be described as follows~\cite{Pandya2020}:
\begin{enumerate}[(a)]
    \item Input data: the state to be tested $\rho$ and any pure product state $\sigma_0$.
    \item Output data: the closest state found after $n$ successes $\rho_n$, and lists of values of $\DHS{\rho}{\rho_i} = \Tr{(\rho - \rho_i)^2}$, $p_i$ and $\sigma_i$.
\end{enumerate}
\begin{enumerate}
    \setcounter{enumi}{-1}
    \item Calculate the value of $\DHS{\rho}{\sigma_0} \equiv \DHS{\rho}{\rho_0}$ and add it to the list.
    \item Increase the counter of trials $c_t$ by 1. Draw a random pure product state $\sigma_1$, hereafter called the first trial state.
    \item Run a preselection for the trial state by checking a value of a linear functional. If it fails, go back to step 1.
    \item In the case of successful preselection, find the minimum of $\Tr{(\rho - \rho_n)^2}$  with respect to $p_n$, which in the case of the first trial state is $\Tr{(\rho - p_1\rho_0 - (1 - p_1)\sigma_1)^2}$ with respect to $p_1$.
    \item If the minimum occurs for $p_1  \in [0, 1]$, then $\rho_1 = p_1\rho_0 + (1 - p_1)\sigma_1$ is saved to the list as the new \ac{CSS}, along with the value of $\DHS{\rho}{\rho_1}$, $p_1$ and $\sigma_1$, increase the success $c_s$ counter value by 1.
    \item Go to step 1., now drawing a new random pure product state $\sigma_2$, such that after a successful preselection, one determines a new $p_2$, and consequently $\rho_2$, and then saves $\DHS{\rho}{\rho_2}$, $p_2$ and $\sigma_2$. Continue in this fashion until a specified HALT criterion is met.
\end{enumerate}

\text{} 

\begin{figure}[t]
    \centering
    \includegraphics[width=0.4\textwidth]{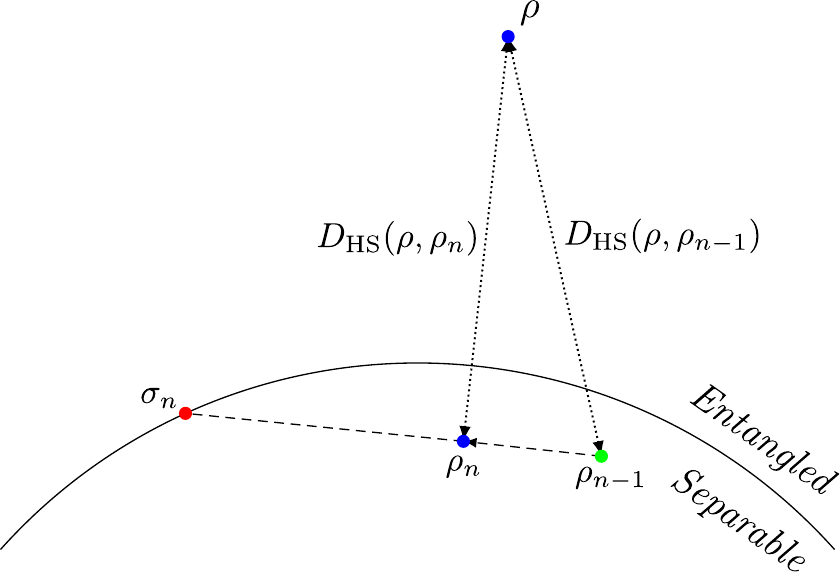}
    \caption{Visualization of the $n^\text{th}$ iteration of the \ac{QGA}. A random pure product state $\sigma_n$ is generated that satisfies the preselection criterion. The next step is to find $\rho_n = p_n\rho_{n-1} + (1 - p_n)\sigma_n$, $p_n \in (0, 1)$, such that $\DHS{\rho}{\rho_n} < \DHS{\rho}{\rho_{n-1}}$. Figure adapted from Ref.~\cite{Pandya2020}.}
    \label{fig:iteration}
\end{figure}

Fig. \ref{fig:iteration} provides a visual representation of the $n^\text{th}$ iteration of the algorithm presented above, while the preselection criterion for the $n^\text{th}$ trial state is given by
\begin{equation}
    \Tr{(\sigma_n - \rho_{n-1})(\rho - \rho_{n-1})} > 0,
    \label{eq:preselection}
\end{equation}
with the geometrical interpretation being that the angle between the vectors $\sigma_n - \rho_{n-1}$ and $\rho - \rho_{n-1}$ is not larger than $\frac{\pi}{2}$~\cite{Pandya2020}, ensuring that a state between $\rho_{n-1}$ and $\sigma_n$ exists such that it is closer to $\rho$. Eq.~\eqref{eq:preselection} can rewritten and computed as
\begin{equation}
    \Tr{\rho_{n-1}^2} + \Tr{\rho\sigma_n} > \Tr{\rho\rho_{n-1}} + \Tr{\rho_{n-1}\sigma_n}.
\end{equation}
The purity of each new \ac{CSS} when calculated iteratively (starting at 1 due to the initial trial state $\sigma_0$ being a pure state), is given by
\begin{align}
    \Tr{\rho_n^2} &= \Tr{(p_n\rho_{n-1} + (1 - p_n)\sigma_n)^2} \nonumber \\ 
    &= p_n^2\Tr{\rho_{n-1}^2} + (1 - p_n)^2 \nonumber \\
    & \hspace{0.5cm} + 2p(1 - p_n)\Tr{\rho_{n-1}\sigma_n},
    \label{eq:qga_purity}
\end{align}
while the overlap is calculated as
\begin{equation}
    \Tr{\rho\rho_n} = p_n\Tr{\rho\rho_{n-1}} + (1 - p_n)\Tr{\rho\sigma_n}.
    \label{eq:qga_overlap}
\end{equation}
Thus, the cost function for the $n^\text{th}$ iteration is defined as follows:
\begin{align}
    \DHS{\rho}{\rho_n} &= \Tr{(\rho - p_n\rho_{n-1} - (1 - p_n)\sigma_n)^2} \nonumber \\
    &= \Tr{\rho^2} + p_n^2\Tr{\rho_{n-1}^2} + (1 - p_n)^2 \nonumber \\ & \hspace{0.5cm} + 2p_n(1 - p_n)\Tr{\rho_{n-1}\sigma_n} \nonumber \\ & \hspace{0.5cm} - 2p_n\Tr{\rho\rho_{n-1}} - 2(1 - p_n)\Tr{\rho\sigma_n},
    \label{eq:cost}
\end{align}
which is explicitly written and computed as
\begin{widetext}
\begin{align}
    \DHS{\rho}{\rho_n} &= \DHS{\rho}{\sum_{i=0}^{n} \left(\prod_{j=i+1}^{n} p_j\right) \left(1-p_i\right)\sigma_i} \nonumber \\
    &= \Tr{\left(\rho - \sum_{i=0}^{n} \left(\prod_{j=i+1}^{n} p_j\right) \left(1-p_i\right)\sigma_i \right)^2} \nonumber \\
    &= \Tr{\rho^2} + \Tr{\left(\sum_{i=0}^{n} \left(\prod_{j=i+1}^{n} p_j\right) \left(1-p_i\right)\sigma_i \right)^2} - 2\Tr{\rho\sum_{i=0}^{n} \left(\prod_{j=i+1}^{n} p_j\right) \left(1-p_i\right)\sigma_i} \nonumber \\
    &= \Tr{\rho^2} + 2\sum_{i < k}^{n}\left(\prod_{j=i+1}^{n} p_j\right)\left(\prod_{l=k+1}^{n} p_l\right) \left(1-p_i\right) \left(1-p_k\right) \Tr{\sigma_i\sigma_k} \ + \nonumber \\ & \hspace{0.5cm} \sum_{i=0}^{n}\left(\prod_{j=i+1}^{n} p_j^2\right) \left(1-p_i\right)^2 - 2\sum_{i=0}^{n} \left(\prod_{j=i+1}^{n} p_j\right) \left(1-p_i\right)\Tr{\rho\sigma_i}.
\end{align}
\end{widetext}
This entails that the overlap between every new trial state $\sigma_n$ and every previous trial state $\sigma_0, \dots, \sigma_{n-1}$, as well as with the test state $\rho$, must be calculated at every iteration. This implies that the parameters of each separable pure trial state must be saved at each success, resulting in an ever-increasing memory and quantum device utilization.

\bibliographystyle{apsrev4-2}
\bibliography{ref.bib}

\end{document}